\documentclass[apjl]{emulateapj}\usepackage{apjfonts}

\begin{document}

\shorttitle{Excited-state OH in AU Gem and NML Cyg}
\shortauthors{Sjouwerman et al.}

\title{Excited-state OH Mainline Masers in AU Geminorum and NML Cygni}
\author{
Lor\'{a}nt~O.~Sjouwerman\altaffilmark{1}, 
Vincent~L.~Fish\altaffilmark{1,2},
Mark~J.~Claussen\altaffilmark{1}, 
Ylva~M.~Pihlstr\"{o}m\altaffilmark{3}, 
\&~Laura~K.~Zschaechner\altaffilmark{3}
}
\altaffiltext{1}{National Radio Astronomy Observatory, 1003 Lopezville
  Rd., Socorro, NM 87801, lsjouwer@nrao.edu, vfish@nrao.edu, mclausse@nrao.edu.}
\altaffiltext{2}{Jansky Fellow.}
\altaffiltext{3}{Department of Physics and Astronomy, University of
  New Mexico, 800 Yale Boulevard NE, Albuquerque, NM 87131, ylva@unm.edu.}

\begin{abstract}

Excited-state OH maser emission has previously been reported in the
circumstellar envelopes of \emph{only two} evolved stars: the Mira
star AU Geminorum and the hypergiant NML Cygni. We present Very Large
Array (VLA) observations of the 1665, 1667, and excited-state 4750~MHz
mainline OH transitions in AU Gem and Expanded Very Large Array (EVLA)
observations of the excited-state 6030 and 6035~MHz OH mainline
transitions in NML Cyg. We detect masers in both mainline transitions
in AU Gem but no excited-state emission in either star. We conclude
that the excited-state OH emission in AU Gem is either a transient
phenomenon (such as for NML Cyg outlined below), or possibly an
artifact in the data, and that the excited state OH emission in NML
Cyg was generated by an episode of enhanced shock between the stellar
mass-loss and an outflow of the Cyg OB2 association. With these single
exceptions, it therefore appears that excited-state OH
emission indeed should not be predicted nor observable in evolved
stars as part of their normal structure or evolution.

\end{abstract}

\section{Introduction}

The topic of hydroxyl (OH) maser pumping in the circumstellar
envelopes of evolved stars has received increased focus over recent
years due to observational and theoretical advances.  On the
observational side, detections of far infrared OH lines toward evolved
stars have confirmed the role of radiative pumping routes and allowed
for estimation of pump efficiencies
\citep{sylvester97,neufeld99,he04,he05}.  On the theoretical side, the
recent \citet{gray05} model demonstrates the complexity of OH pumping
routes and is quite successful at explaining observed OH maser
properties from OH/IR stars. Pump models generally do not predict
detectable excited-state emission in circumstellar envelopes of
evolved stars, either due to a lack of inversion or insufficient
optical depth (e.g., \citealt{elitzur76}; \citealt{bujarrabal80};
M.~D.\ Gray 2006, private communication). For a comprehensive overview
on circumstellar shells around evolved stars we refer to
\citet{habing96}.

Excited-state OH masers have been sought in the 4.8 and 6.0~GHz
transitions toward the circumstellar environments of a variety of
evolved high mass losing stellar sources by many authors
\citep{thacker70,zuckerman72,baudry74,rickard75,claussen81,jewell85,desmurs02,fish06}.
These searches universally failed to detect excited-state emission
except in two circumstellar environments.  \citet{claussen81} report
on a $5\,\sigma$ detection of a 4750~MHz maser in \object{AU Gem}, a
Mira variable.  Subsequent observations by \citet{jewell85} failed to
confirm this emission, but the RMS noise level of their observations
does not rule out the 100~mJy detection of \citet{claussen81}.

The second evolved star with an excited-state maser detection
is \object{NML Cyg}, a supergiant and recently more often referred to
as a hypergiant \citep{vgenderen82,schuster06}.  \citet{zuckerman72}
report a clear ($\gg 10\,\sigma$) detection at 6035~MHz and possible
emission ($\sim3\,\sigma$) at 6030~MHz as well.  Their spectra
indicate strong (2.2~Jy) 6035~MHz maser emission at $V_\mathrm{LSR}
\approx 4$~km\,s$^{-1}$. This is close to the stellar velocity of
about 0 ($\pm$ 4) km s$^{-1}$ \citep{zhou81,diamond84}, usually found
as the average velocity of the outer boundaries or emission peaks of
the characteristic double-peaked 1612 MHz OH emission or as the center
of the SiO emission in OH/IR stars \citep{bowers83,habing96}.  With
ample sensitivity to follow up on this initial detection,
\citet{jewell85} did not detect any emission from NML Cyg a decade
later.  In 1999, \citet{desmurs02} also did not detect the emission at
4~km\,s$^{-1}$ but note possible weak ($\sim 20$~mJy) emission at
$-17$~km\,s$^{-1}$, corresponding in velocity to an inner 1612~MHz
peak near $-18$~km\,s$^{-1}$ \citep[e.g.,][]{herbig74,engels79}.  The
original spectra of \citet{zuckerman72} are suggestive of weak
emission at both 6030 and 6035~MHz near this velocity, but not at
significant levels compared to their RMS noise.

The confirmed detection of excited-state OH masers in the
circumstellar environments or ejecta of evolved stars would provide a
challenge to modern pumping models, or at least highlight novel
portions of pumping phase space heretofore not considered.  To date,
neither report of a detection of an excited-state OH maser in an
evolved star has been confirmed, which in view of the pumping models
therefore needs to be followed up.  In this Letter, we present new
observations of the OH masers in AU~Gem and NML~Cyg in the
excited-state OH lines originally reported as detections, as well as
ground-state 1665 and 1667 MHz mainline OH emission in AU Gem.

\section{Observations}

\begin{table}\label{tab:obs}
\caption{Observational summary table}\vspace*{-0.7cm}
\begin{center}\begin{tabular}{lccccc}
Observing Date & Line  & $\Delta \nu$ & $\Delta V$    & Beam                            & RMS \\
     & (MHz) & (kHz)        & (km s$^{-1}$) & ($\arcsec\times\arcsec, \degr$) &  \tablenotemark{\dag} \\
\noalign{\smallskip}\tableline\noalign{\smallskip}
\tableline\noalign{\smallskip}
\multicolumn{6}{l}{\ \ \ AU Gem, centered at $V_\mathrm{LSR} \approx$ 10 km s$^{-1}$} \\
\noalign{\smallskip}
 2007 Jan 25     & 1665  & $\phantom{0}$195 & 0.28 & 40.3$\times$21.5, 73  & 4.9\tablenotemark{a} \\
 2007 Jan 25     & 1667  & $\phantom{0}$195 & 0.28 & 40.4$\times$20.4, 72  & 4.7\tablenotemark{a} \\
 2007 Jan 25     & 4750  & $\phantom{0}$781 & 0.19 & 13.0$\times$12.4, 94  & 3.3\tablenotemark{b} \\
\noalign{\smallskip}
\multicolumn{6}{l}{\ \ \ NML Cyg, centered at $V_\mathrm{LSR} \approx$ 0 km s$^{-1}$} \\
\noalign{\smallskip}
 2007 May 26/30  & 6030  &             1563 & 0.61 & 0.56$\times$0.23, 118 & 8.2\tablenotemark{c} \\
 2007 May 26/30  & 6035  &             1563 & 0.61 & 0.58$\times$0.23, 118 & 7.1\tablenotemark{d} \\
\noalign{\smallskip}\tableline
\end{tabular}\end{center}
\tablenotetext{\dag}{Typical RMS noise in a spectral channel in mJy beam$^{-1}$}
\tablenotetext{a}{Assuming a flux density of 2.0 Jy at 1.7 GHz for J0741+312}
\tablenotetext{b}{Assuming a flux density of 1.6 Jy at 4.8 GHz for J0741+312}
\tablenotetext{c}{With a measured flux density of 2.67($\pm$0.10) Jy for J2052+365}
\tablenotetext{d}{With a measured flux density of 2.62($\pm$0.07) Jy for J2052+365}
\end{table}

AU Gem and NML Cyg were observed with the Very Large Array (VLA) using
the settings as outlined in Table~1.
 The AU Gem observations occurred in DnC
configuration.  Due to the Expanded VLA (EVLA) upgrade
\citep{mckinnon01,ulvestad07}, only 22 antennas were available,
including three EVLA antennas. All these antennas were used.

For NML Cyg we were able to profit from the special call for proposals
in 2007 April for using the new 5 GHz (C-band) receivers on the EVLA.
These receivers are part of the EVLA upgrade and allow observing at a
much wider frequency range, in our case at 6030 and 6035 MHz.
With the configuration in A-array,
three EVLA antennas were spread near-homogeneously over the North and
West arm each. Though the East arm had a similar distribution of EVLA
antennas, the outer two were not operational at 6.0 GHz.

For AU Gem, two hours were devoted to the 4750~MHz line followed by
two hours on the 1665 and 1667~MHz lines of OH. Due to the lack of
observations of an absolute standard calibrator (e.g., 3C286), the
source 0741+312 was used for bandpass, phase and amplitude, and
primary flux calibrations, using assumed flux densities of 2.0~Jy at
1.7~GHz and 1.6~Jy at 4.8~GHz as taken from the VLA calibrator list.
System flux monitoring suggests this is in error with less than 20\%.
As a temporary inconvenience, online Doppler tracking was not used due
to differences in the VLA and EVLA antenna control systems.  Instead,
the sky frequency was calculated from the LSR velocity using the NRAO
online Dopset
tool and
the observations were taken in fixed-frequency mode.  The LSR velocity
scale is set to the mean sky frequency of the observations.  Maximum
deviations from this frequency (at the beginning and end of observing
in each band) correspond to 0.50 channel widths at 4.8~GHz ($\lesssim$
0.1 km s$^{-1}$) and 0.36 channel widths at 1.7~GHz ($\lesssim$ 0.1 km
s$^{-1}$).

Two 1.5 hour blocks were devoted to observations of the 6030 and
6035~MHz lines in NML Cyg.  By using only the EVLA antennas, online
Doppler tracking was available.  The data were calibrated in the
standard way for VLA antennas using 3C48 as standard flux calibrator
and J2052$+$365 as phase calibrator. The new EVLA system required more
careful flagging of bad data than usual with the established VLA
system. Data from some antennas were discarded entirely: one antenna
did not have phase stability, while another antenna had no signal in
the 6030 MHz LCP IF. A third antenna showed occasional phase jumps of
about 60$\degr$. The absolute flux calibration at 6.0 GHz has not yet
been determined; flux densities and RMS noise levels given in
Table~1
 assume a scaling with frequency for 3C48 as
calculated in the AIPS task SETJY \citep{greisen03}.

\section{Results}

\begin{figure}[t]
\plotone{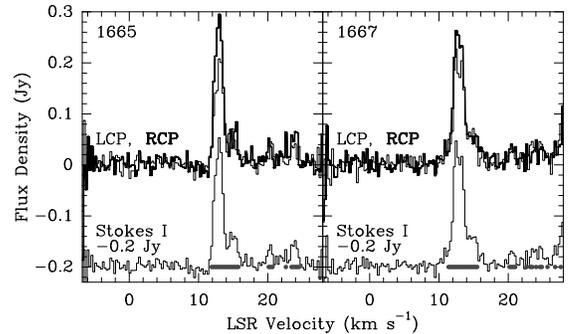}
\caption{Left: Spectra of the 1665~MHz emission in AU~Gem.  RCP
  emission is shown in bold and LCP in normal weight.  The Stokes I
  spectrum has been shifted vertically for clarity.  Grey dots
  indicate channels in which emission is detected at the phase center at
  more than 5 times the RMS noise. Right: Spectra of the 1667~MHz
  emission in AU~Gem.
  \label{fig-1665}
  }
\end{figure}

\subsection{AU Gem}\label{sect:aug}

Figure~\ref{fig-1665} 
shows the emission at 1665 and 1667~MHz
in AU Gem.  We detect $\sim 250$~mJy emission in each of the 1665 and
1667~MHz transitions in the 12--13~km\,s$^{-1}$ range, corresponding
to similarly bright 1667~MHz emission detected by \citet{rieu79}.
Right circular polarized (RCP) emission is stronger in each transition
than left circular polarized (LCP) emission.  The second peak near
15~km\,s$^{-1}$ corresponds to the broader redshifted shoulder of the
emission in the lower-resolution \citet{rieu79} spectrum.  We detect
weaker features at higher LSR velocity as well, including emission in
the edge spectral channels at $\sim 28$~km\,s$^{-1}$ at 1667~MHz, but
we do not detect the weaker ($\sim 70$~mJy) emission at 0~km\,s$^{-1}$
claimed by \citet{rieu79}.  If we assume that the 1665 and 1667 MHz
mainline OH-peaks at the systemic velocity of 13 km\,s$^{-1}$ define
the stellar velocity, and if we conclude that the \citet{rieu79}
$\sim$ 0 km\,s$^{-1}$ 1667~MHz weak feature is convincing, then these
features could indicate an expanding OH shell with an expansion
velocity of about 14-15 km\,s$^{-1}$. If we were to dismiss the
\citet{rieu79} $\sim$ 0 km\,s$^{-1}$ 1667~MHz feature, the systemic
and expansion velocities would be 20 and 8 km\,s$^{-1}$,
respectively. Both expansion velocities are reasonable for Mira-type
stars with the latter for a less optically thick (lower metallicity)
shell \citep{habing96}, but the uncertainties in the spectra do not
allow a firm conclusion on this. In this respect it is unfortunate
that AU Gem has never been detected in the 1612 MHz OH transition
\citep{fix78,rieu79,olnon80} nor shows any H$_2$O nor SiO emission
\citep{nyman86}. All emission is consistent with being
point-like at the 40\arcsec$\times$20\arcsec\ resolution of the VLA in
DnC configuration, which is as expected since a typical Mira OH shell
size \citep[$10^{16}$~cm, from][]{herman85} would subtend an angle of
less than 0\farcs3 at the distance of AU Gem \citep[2.4 kpc,
from][]{rieu79}.

As to the 4750 MHz transition, we do not detect any emission at a
$3\,\sigma$ noise level of 10~mJy beam$^{-1}$ in AU Gem.

\subsection{NML Cyg}

\begin{figure}[t]
\plotone{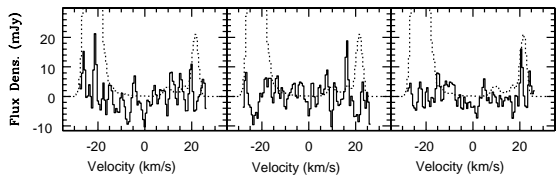}
\caption{Hanning smoothed spectra of tentative 6035~MHz emission
  within 1\arcsec\ of the position of NML~Cyg. Arbitrarily scaled
  1612~MHz emission from VLA archival data is shown with a dotted line
  for reference.
  \label{fig-6035}
  }
\end{figure}

No emission was found over 4.2$\sigma$ (34.2 mJy beam$^{-1}$) in the
6030 MHz data, nor was any emission found over 4.9$\sigma$ (35.3 mJy
beam$^{-1}$) in the 6035 MHz data within the beam of the
\citet{jewell85} observations.  We cannot rule out tentative
($>$4.8$\sigma$ single channel) 6035 MHz line emission at $-$21.2,
$+$16.4 and $+$20.6 km s$^{-1}$ (Fig.~\ref{fig-6035}). This would
support the notion of tentative 6035 MHz OH features at $\approx -$17
km s$^{-1}$ \citep{desmurs02,zuckerman72} and could be related to a
new episode of high mass loss or compression of circumstellar material
close to the star as traced by the collisionally excited SiO maser
\citep{boboltz00}. However, we would not claim such emission without
extensive high-sensitivity follow-up observations, e.g., when all VLA
antennas have new C-band receivers installed (in 2010).

\section{Discussion}

Theoretical work on OH pumping in model, well-behaved circumstellar
shells predict that excited-state OH masers should not be seen, and
observations in most circumstellar shells to date agree. Our
non-detections of excited-state OH in AU Gem in the 4750~MHz line and
in NML Cyg in the 6030 and 6035 MHz lines support the prediction that
excited-state OH should not be observable in circumstellar shells as
part of their normal structure and evolution. Excepting temporary
events that cannot easily be reobserved, the previous reports of
detection of the excited-state OH lines in AU Gem and NML Cyg cannot be
explained.

\subsection{AU Gem}

Dismissing the not very sensitive observation by \citet{jewell85}, our
10 mJy beam$^{-1}$ $3\sigma$ upper limit non-detection of the 4750~MHz
excited-state OH emission contrasts the sole report of a $\sim$100 mJy
$5\sigma$ positive detection in this line by \citet{claussen81}. If
not an unrelated source in the line-of-sight of the 0$\farcm$5 beam
of the Arecibo telescope, this suggests two possibilities.

The first is that the 4750~MHz maser emission in AU Gem has weakened
substantially (by more than a factor of 10) between 1981 and 2007.
The lack of other known evolved stars with 4750~MHz OH masers
precludes us from commenting on the phenomenology in this transition,
but it is worth noting that in a comparable time period the 6035~MHz
line in NML Cyg weakened by more than a factor of 100 \citep[][and
discussed below]{zuckerman72,desmurs02}.  The second possibility is
that the \citet{claussen81} detection is spurious.  It is based on a
$5\,\sigma$ spike in a single 0.6~km\,s$^{-1}$ spectral channel in
frequency switched Arecibo data.  It is possible that this emission
arises from terrestrial interference or a bad correlator channel
rather than a celestial source.  The LSR velocity of this feature,
3.5~km\,s$^{-1}$, is outside the range of the dominant emission in
either ground-state main line.  However, it would still fall within
the derived velocity range of the outer 1667 MHz OH-peaks as argued in
the first ($V_{\rm exp} = 14-15$ km s$^{-1}$) case in
Sect.~\ref{sect:aug}.  Even though the 1612 MHz OH, H$_2$O and SiO
transitions have not been detected
\citep{fix78,rieu79,olnon80,nyman86}, making AU Gem atypical for an
(Type II) OH/IR star, we could not argue a special case scenario for
AU Gem as for NML Cyg outlined below. AU Gem probably just does not
have a circumstellar environment thick or dusty enough to sustain 1612
MHz maser emission and thus we think that it is unlikely to have ever
had excited-state OH emission.

\subsection{NML Cyg}\label{sect:nml}

The new non-detections of 6030 and 6035 MHz emission toward NML Cyg
suggest two possibilities.  Like AU Gem, this phenomenon could be a
time-variable or single event in the history of NML Cyg as perhaps
typical for any other OH/IR object during their evolution. However, as
stated before, because other searches for excited-state OH emission in
circumstellar environments of a variety of evolved stars have yielded
no detections, it appears that NML Cyg must somehow be special to have
had 6.0 GHz excited-state OH maser emission. We reject the possibility
of the original multi-channel $\gg10\sigma$ detection by
\citet{zuckerman72} as being spurious. The second possible explanation
is that NML Cyg \emph{indeed is special} compared to other evolved
OH/IR type stars due to the interaction with its environment.

Since the discovery of emission in the ground-state OH lines in the
optically obscured infrared star NML Cyg by
\citet{wilbar68a,wilbar68b}, it has been extensively studied in the
radio mainly in the very bright 1612 MHz satellite line of OH.
  The main lines at 1665 and 1667 MHz are present as
well \citep[][ and from VLA archive data]{wilbar68a,wilbar68b}, but
with typically a factor of 100 less flux than the 1612 MHz emission,
whereas the 1720 MHz satellite line was reported to be undetected by
\citet{wilbar68a,wilbar68b}.  For comparison in the following
discussion we refer to the high resolution (MERLIN) ground-state
blueshifted 1665 MHz mainline OH and full extent 1612 MHz OH data on
NML Cyg described by \citet{diamond84} and \citet{etoka04}.

Though the asymmetric morphology in the 1612 MHz OH transition was
modeled at first as a rotating disk at a position angle of about
150$\degr$ toward the northwest \citep[e.g.,][]{masheder74,benson79},
subsequent interpretations favor a double 1612 MHz OH circumstellar
shell \citep{diamond84,etoka04} with the inner shell as a spherical
expanding shell typical for an OH/IR star.  The outer 1612 MHz OH
shell only manifests itself as a curved arc-like
shell-segment located to the northwest at about 2\farcs3 from the star
at a velocity close to the stellar velocity, where the shell motion is
predominantly tangential in the plane of the sky \citep[][their
Figures 8 and 9]{etoka04}. The incomplete (blueshifted emission only)
1665 MHz mainline OH image shows a similar picture, with a less
prominent arc located on the sky at about 1\farcs3 to the
northwest and between the two 1612 MHz shells \citep[][their Figure
7]{etoka04}.  It is noteworthy that the mass outflow is not
constant; e.g., \citet{danchi01} deduce a 3.86 mas yr$^{-1}$ infrared
proper motion of another double shell closer ($<$ 0\farcs3, versus
2\farcs3 for the 1612 MHz OH arc) to the star, and an age difference
between the two inner expanding shells of 65 ($\pm$14) years.

Toward the west-northwest, NML Cyg is surrounded by an \ion{H}{2}
region \citep{habing82}, originating from the ionization of outflowing
material from NML Cyg by the intense photoionizing radiation field of
the nearby Cyg OB2 association \citep{morris83}. It is in particular
interesting to note that \citet{habing82} overlay their \ion{H}{2}
observations on the red Palomar Sky Atlas {\it E} plate and identify a
near-linear feature at about 30-35\arcsec\ west-northwest from NML
Cyg. \citet{habing82} suggest that this depicts H$\alpha$ emission,
which we in turn recognize as a tracer for shocked material. Recent
HST observations by \citet{schuster06} of the dust immediately
surrounding NML Cyg outlines the dissociation surface generated by
this radiation field, oriented in the same direction as the \ion{H}{2}
region and the 1612 MHz OH arc (but with 0\farcs25 extent at a much
smaller scale). We find it plausible that a bow shock-like front
causes the OH molecules and dust toward the direction of the Cyg OB2
association to pile up, building up sufficient OH (column) density.
The increased absorption of stellar radiation by the enhanced density
of dust will radiatively pump the 1612 MHz maser, causing a partial
shell or arc to appear at a line-of-sight velocity near the stellar
velocity \citep{etoka04}.

We note that 1720 MHz satellite OH masers generally are seen in high
density shocks, as is H$\alpha$ emission. However, recent calculations
\citep{wardle07} show that excited-state OH might be observable at
even higher densities.  The variation in mass loss of NML Cyg may
cause the northwest side pile up to occasionally be shocked and
temporarily have increased density due to shells impacting on the
near-stationary material. It could have been that \citet{zuckerman72}
observed NML Cyg during such an impact event. Since then the
shock likely has dissipated and the densities have fallen. If the 65
years between the shells found by \citet{danchi01} is typical,
currently we are observing halfway between the impact of two shells,
predicting another impact around the years 2030-2040. We speculate
that while the excited-state emission is now gone, ground-state 1720
MHz OH emission may be detectable if the pumping is predominantly
collisional.  Observations of the 1720 MHz transition at greater
sensitivity ($\ll$ 1 Jy) than that obtained by
\citet{wilbar68a,wilbar68b} are required to test this hypothesis.

Future discussions of maser pumping models should therefore not be
distracted by these two cases of reported excited-state emission in
circumstellar environments of late-type stars anymore. Instead,
current models seem sufficient to explain the
pumping of masers in circumstellar shells of evolved stars as part of
their normal structure and evolution.

\section{Summary}

Although a temporary event in AU Gem causing 4750 MHz excited-state OH
emission cannot be excluded, we conclude that it is probable that the
original report by \citet{claussen81} regards an unfortunate spurious
detection and is not 4750 MHz excited-state OH maser emission in the
circumstellar environment of AU Gem.

The reported 6035 MHz excited-state OH emission by \citet{zuckerman72}
in NML Cyg cannot be spurious and can be explained as originating from
the special temporary conditions arising from a shock between the high
mass outflow of NML Cyg and the intense ionizing UV-radiation field of
the nearby Cyg OB2 association. Although it is clear that the 6035 MHz
emission in NML Cyg has weakened substantially, it is unclear whether
the excited-state OH emission is still present at very low levels due
to a lack of sensitivity in this early-stage of the VLA to EVLA
transition. We suggest that the excited-state OH transitions in NML
Cyg be reobserved with higher sensitivity when there are more EVLA
antennas with this capability. We also recommend that the 1720 MHz OH
transition be observed as a tracer of a possible shock in the
environment of NML Cyg.

The transition from the VLA to the EVLA is well underway and already
now offers great opportunities to observe at frequencies previously
unavailable to the VLA user community.

\acknowledgments

The National Radio Astronomy Observatory is a facility of the National
Science Foundation operated under cooperative agreement by Associated
Universities, Inc.

{\it Facilities: \facility{VLA, EVLA}}

\end{document}